\documentclass{ws-procs9x6}

\newcommand{\bra}{\langle}

\begin{document}

\title{$B\to\pi,K,\eta$ Decay Formfactors from
    Light-Cone Sum Rules}

\author{P. BALL}

\address{Department of Physics\\
University of Durham \\ 
Durham DH1 3LE, UK\\ 
E-mail: patricia.ball@durham.ac.uk}

\author{R. ZWICKY}

\address{William I. Fine Theoretical Physics Institute\\ 
University of Minnesota\\
Minneapolis, MN 55455, USA\\
E-mail: zwicky@physics.umn.edu}  

\maketitle

\abstracts{
We present an improved calculation of all $B\to$ light pseudoscalar 
formfactors from light-cone sum rules, including one-loop radiative
corrections to twist-2 and twist-3 contributions, and leading order
twist-4 corrections. The total theoretical uncertainty of our
results at zero momentum transfer is 10 to 13\%. The dependence of the
formfactors on the momentum transfer $q^2$ is parametrized in a simple way
that is consistent with their analytical properties and is valid for
all physical $q^2$. The uncertainty of the extrapolation in $q^2$
on the semileptonic decay rate $\Gamma(B\to\pi e \nu)$ is estimated to
be 5\%.
}

\section{Introduction and Definitions}

In a recent paper\cite{neu} we have reported a new calculation of
$B\to\pi,K,\eta$ decay formfactors 
from QCD sum rules on the light-cone (LCSRs). The
paper improves upon our previous publications\cite{PB98,ballroman} by:
\begin{itemize}
\item including radiative corrections to twist-3 contributions to
  one-loop accuracy, for all formfactors;
\item a precisely defined method for fixing sum rule specific
  parameters;
\item using updated values for input parameters;
\item a careful analysis of the uncertainties of the formfactors at
  zero momentum transfer;
\item a new parametrization of the dependence of the formfactors on
  momentum transfer, which is consistent with the constraints from
analyticity and  heavy-quark expansion;
\item a detailed breakdown of the dependence of formfactors on nonperturbative
  hadronic parameters describing the $\pi$, $K$, $\eta$ mesons, which
  facilitates the incorporation of future
  updates of their numerical values and also allows a consistent treatment
  of their effect on nonleptonic decays.
\end{itemize}
The key idea of LCSRs is to
consider a correlation function of the weak current and a current with
the quantum-numbers of the $B$ meson, sandwiched between the vacuum and,
in the present context, the pseudoscalar meson $P$,
i.e.\ $\pi$, $K$ and $\eta$. 
For large (negative) virtualities of these currents, the
correlation function is, in coordinate-space, dominated by distances
close to the light-cone and can be discussed in the framework of
light-cone expansion. In contrast to the short-distance expansion
employed by conventional QCD sum rules \`a la SVZ\cite{SVZ}, where
nonperturbative effects are encoded in vacuum expectation values 
of local operators with
vacuum quantum numbers, the condensates, LCSRs
rely on the factorisation of the underlying correlation function into
genuinely nonperturbative and universal hadron distribution amplitudes (DAs)
$\phi$ that are convoluted with process-dependent amplitudes $T$,
which are the analogues to the Wilson-coefficients in the
short-distance expansion and can be
calculated in perturbation theory. Schematically, one has
\begin{equation}\label{eq:1}
\mbox{correlation function~}\sim \sum_n T^{(n)}\otimes \phi^{(n)}.
\end{equation}
The sum runs over contributions with increasing twist, labelled by
$n$, which are suppressed by
increasing powers of, roughly speaking, the virtualities of the
involved currents. 
The same correlation function can, on the other hand, be written as a
dispersion-relation, in the virtuality of the current coupling to the
$B$ meson. Equating dispersion-representation and the
light-cone expansion, and separating the $B$ meson contribution from
that of higher one- and multi-particle states, one obtains a relation
(QCD sum rule) for the formfactor describing the $B\to P$ transition. 

The particular strength of LCSRs lies in the
fact that they allow the inclusion not only
of hard-gluon exchange contributions, which have been identified, in
the seminal papers that opened the study of hard exclusive processes
in the framework of perturbative QCD (pQCD)\cite{pQCD}, 
as being dominant in light-meson form
factors, but that they also capture the so-called
Feynman-mechanism, where the quark created at the weak vertex carries
nearly all momentum of the meson in the final state, while
all other quarks are soft. This mechanism is suppressed by two powers
of momentum-transfer in processes with light mesons, but there 
is no suppression in heavy-to-light
transitions\cite{CZ}, and hence any reasonable 
application of pQCD to $B$ meson
decays should include this mechanism. 
It is precisely LCSRs that
accomplish this task and have been applied to a
variety of problems in heavy-meson physics.\cite{neu,PB98,ballroman,LCSRs}
A more detailed discussion of the
rationale of LCSRs and of the more
technical aspects of the method can 
be found e.g.\ in Ref.\cite{LCSRs:reviews}. 

The formfactors in question can be defined as ($q=p_B-p$)
\begin{eqnarray}
\langle P(p) | \bar q \gamma_\mu b | B(p_B)\rangle &=&  
f_+^P(q^2) \left\{(p_B+p)_\mu - \frac{m_B^2-m_P^2}{q^2} \, q_\mu 
\right\}\nonumber\\
&&{} +
\frac{m_B^2-m_P^2}{q^2} \, f_0^P(q^2)\, q_\mu,\label{FF1}\\
\langle P(p) | \bar q \sigma_{\mu\nu} q^\nu b | B(p_B)
\rangle 
&=&i\left\{ (p_B+p)_\mu q^2 - q_\mu (m_B^2-m_P^2)\right\} \,
  \frac{f_T^P(q^2)}{m_B+m_P}\,.\quad
\end{eqnarray}
The starting point for the calculation of e.g.\ $f_+^\pi$  is the
correlation function
\begin{equation}
i\int d^4y e^{iqy} \langle \pi(p)|T[\bar q\gamma_\mu b](y)
[m_b\bar b i\gamma_5 q](0)|0\rangle
=\Pi_+ 2p_\mu + \dots,\label{eq:CF}
\end{equation}
where the dots stand for other Lorentz structures. 
For a certain configuration of
virtualities, namely $m_b^2-p_B^2\geq O(\Lambda_{\rm QCD}m_b)$ and 
$m_b^2-q^2\geq
O(\Lambda_{\rm QCD}m_b)$, the integral is dominated by light-like distances 
and can be expanded around the light-cone:
\begin{equation}\label{eq:3}
\Pi_+ (q^2,p_B^2) = \sum_n \int_0^1 du\, \phi^{(n)}(u;\mu_{\rm F}) 
T^{(n)}(u;q^2,p_B^2;\mu_{\rm F}).
\end{equation}
As in Eq.~(\ref{eq:1}), $n$ labels the twist of operators and 
$\mu_{\rm F}$ denotes the factorisation scale. The restriction
on $q^2$, $m_b^2-q^2\geq O(\Lambda_{\rm QCD}m_b)$, 
implies that $f_+^\pi$ is not accessible at all momentum-transfers; to
be specific, we restrict ourselves to $0\leq q^2\leq 14\,$GeV$^2$.
As $\Pi_+$ is independent
of $\mu_{\rm F}$, the above formula implies that the scale-dependence of
$T^{(n)}$ must be canceled by that of the DAs $\phi^{(n)}$. 

In Eq.~(\ref{eq:3}) it is assumed that $\Pi_+$ can be described by
collinear factorisation, i.e.\ that the only relevant degrees of
freedom are the longitudinal momentum fractions $u$ carried by the
partons in the $\pi$, and that
transverse momenta can be integrated over. Hard infrared (collinear) 
divergences occurring in $T^{(n)}$ should be absorbable into the
DAs. Collinear factorisation is trivial at tree-level,
where the $b$ quark mass acts effectively as regulator,
but can, in principle, be violated by radiative corrections, by
so-called ``soft'' divergent terms, which yield divergences upon
integration over $u$. Actually, however, it turns out that for all
formfactors calculated in Ref.\cite{neu} the
$T$ are nonsingular at the endpoints $u=0,1$, so there are {\em no soft
divergences, independent of the end-point behavior of the distribution
amplitudes.} In Ref.\cite{neu} Eq.~(\ref{eq:3}) has been demonstrated to be 
valid to $O(\alpha_s)$ accuracy
for twist-2 and twist-3 contributions for all correlation functions
$\Pi_{+,0,T}$ from which to determine the formfactors $f_{+,0,T}$.

As for the distribution amplitudes (DAs), they have been discussed
intensively in the literature\cite{DAs}. For pseudoscalar mesons, 
there is only one DA of leading-twist, i.e.\ twist-2, which is 
defined by the following light-cone matrix element ($x^2=0$):
\begin{equation}
\langle 0 | \bar u(x)\gamma_\mu\gamma_5 d(-x) |
 \pi(p)\rangle = i
f_\pi p_\mu \int_0^1 du e^{i\zeta px} \phi_\pi(u),
\end{equation}
where $\zeta=2u-1$ and we have suppressed the Wilson-line $[x,-x]$
 needed to ensure
gauge-invariance. The higher-twist DAs are of type
$$\bra 0 | \bar u(x) \Gamma d(-x)|\pi(p)\rangle \quad \mbox{or}\quad
\bra 0 | \bar u(x)\Gamma G^a_{\mu\nu}(vx)\lambda^a/2  
d(-x)|\pi(p)\rangle,
$$
where $v$ is a number between 0 and 1 and $\Gamma$
a combination of Dirac matrices.
The sum rule calculations performed in 
Refs.\cite{neu,PB98,ballroman} include
all contributions from DAs up to twist-4. The DAs are parametrized by
 their partial wave expansion in conformal spin, which to NLO provides a
 controlled and economic expansion in terms of only a few hadronic
 parameters\cite{DAs}.

The LCSR for $f_+^\pi$ is derived in the following way: the
correlation function $\Pi_+$, calculated for unphysical
$p_B^2$, can be written as dispersion relation over its physical cut. Singling
out the contribution of the $B$ meson, one has
\begin{equation}\label{eq:corr}
\Pi_+ =  f_+^\pi(q^2) \, \frac{m_B^2f_B}{m_B^2-p_B^2}
+ \mbox{\rm higher poles and cuts},
\end{equation}
where $f_B$ is the leptonic decay constant of the $B$ meson,
$f_Bm_B^2=m_b\langle B| \bar b i\gamma_5 d|0\rangle$.
In the framework of LCSRs one does not use (\ref{eq:corr}) as it stands,
but performs a  Borel transformation,
$1/(t-p_B^2)\to \hat{B}\, 1/(t-p_B^2) = 1/M^2 \exp(-t/M^2)$,
with the Borel parameter $M^2$; this transformation enhances the
ground-state $B$ meson contribution to the dispersion representation of $\Pi_+$
and suppresses contributions of higher twist to the light-cone expansion of
$\Pi_+$. The next step is to invoke quark-hadron
duality to approximate the contributions of hadrons other than the
ground-state $B$ meson by the imaginary part of the light-cone
expansion of $\Pi_+$, so that
\begin{eqnarray}
\hat{B}{\Pi_+^{\rm LCE}} & = &
\frac{1}{M^2}\, m_B^2f_B \,f_+^\pi(q^2)\,e^{-m_B^2/M^2}\nonumber\\
&&{} +
\frac{1}{M^2}\, \frac{1}{\pi}\int_{s_0}^\infty \!\! dt \, {\rm
Im}{\Pi^{\rm LCE}_+}(t) \, \exp(-t/M^2)\,.
\end{eqnarray}
Subtracting the 2nd term on the right-hand side from both sides, one
obtains
\begin{eqnarray}
\frac{1}{M^2}\, \frac{1}{\pi}\int_{m_b^2}^{s_0} \!\! dt \, {\rm
Im}{\Pi^{\rm LCE}_+}(t) \, \exp(-t/M^2) & = & \frac{1}{M^2}\,
m_B^2f_B \,f_+^\pi(q^2)\,e^{-m_B^2/M^2}.\label{eq:SR}
\end{eqnarray}
Eq.~(\ref{eq:SR}) is the LCSR for $f_+^\pi$.
$s_0$ is the so-called continuum
threshold, which separates the ground-state from the continuum
contribution. At tree-level, the continuum-subtraction in
(\ref{eq:SR}) introduces a lower limit of integration in $u$, the
momentum fraction of the quark in the $\pi$: $u\geq
(m_b^2-q^2)/(s_0-q^2)$, in (\ref{eq:3}), which behaves as
$1-\Lambda_{\rm QCD}/m_b$ for
large $m_b$ and thus corresponds to the dynamical
configuration of the Feynman-mechanism, as it cuts off low momenta of
the $u$ quark created at the weak vertex. At $O(\alpha_s)$, there are
also contributions with no cut in the integration over $u$, which
correspond to hard-gluon exchange contributions. 
The task now is to find sets of
parameters $M^2$ (the Borel parameter) and $s_0$ (the continuum
threshold) such that the resulting formfactor does not
depend too much on the precise values of these parameters.

\section{Results}

\begin{figure}[t]
$$
\epsfxsize=0.45\textwidth\epsffile{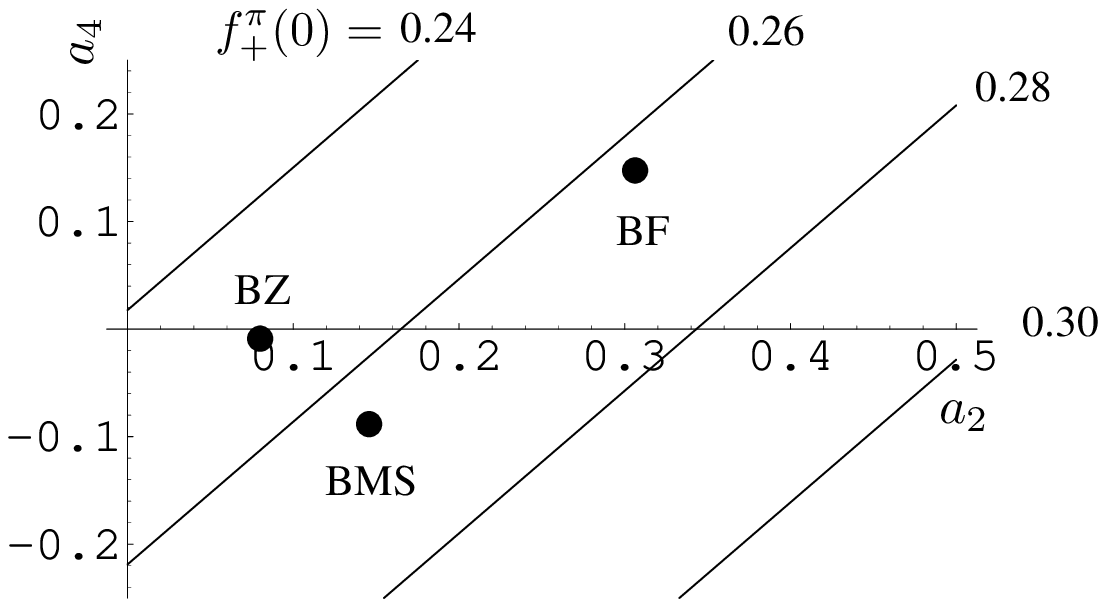}
$$
\vskip-0.4cm
\caption[]{Dependence of $f_+^\pi(0)$ on $a_2$ and
  $a_4$, for central values of input parameters. The lines are lines of
  constant $f_+^\pi(0)$. The dot labeled BZ denotes our preferred
  values 
of $a_{2,4}$, BMS the values from the nonlocal condensate 
model\cite{BMS} and BF from sum rule calculations\cite{DAs}.}\label{fig:1}
$$\epsfxsize=0.45\textwidth\epsffile{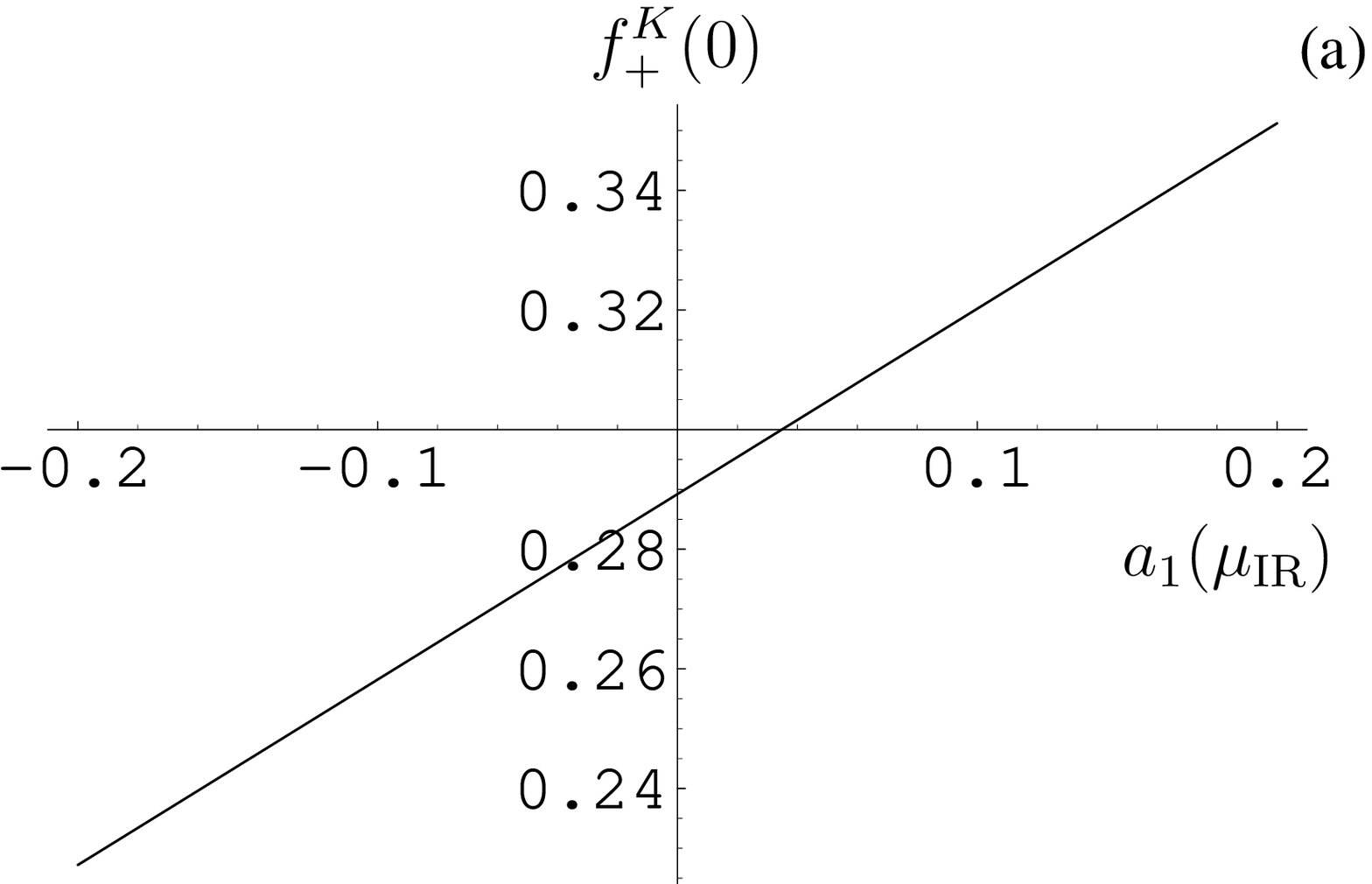}\qquad 
\epsfxsize=0.45\textwidth\epsffile{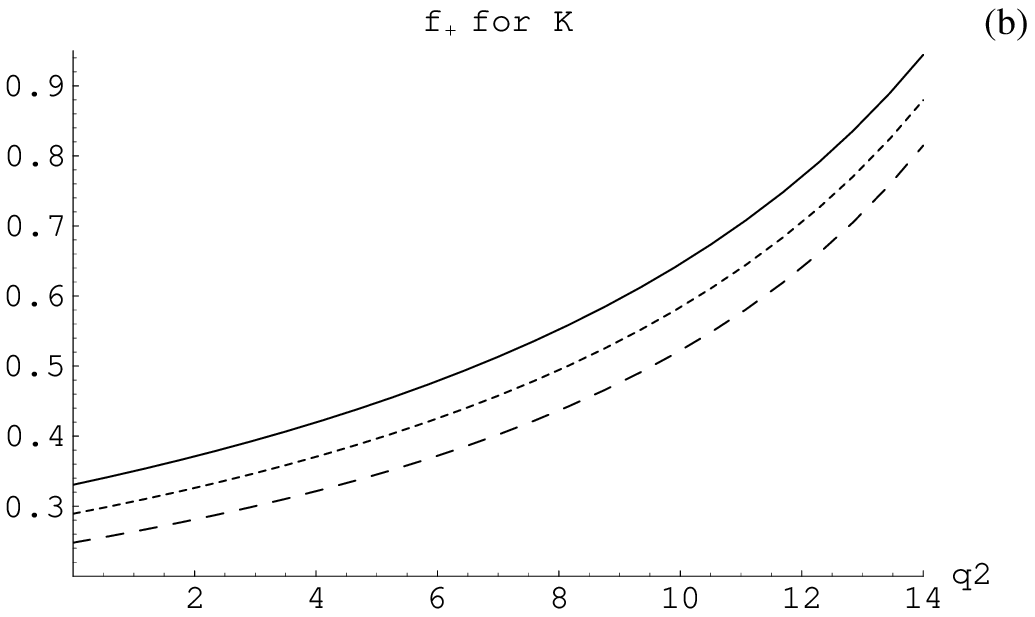}$$
\vskip-0.4cm
\caption[]{(a) Dependence of $f_+^K(0)$ on the Gegenbauer moment
$a_1$. (b) $f_+^K(q^2)$ as function of $q^2$ for
  different values of $a_1$: solid line: $a_1^K = 0.17$,
  short dashes: $a_1^K = 0$,
  long dashes: $a_1^K = -0.18$.}\label{fig:2}
\end{figure}

For a detailed discussion of the procedure used to determine the
hadronic and sum rule specific input parameters we refer to
Ref.\cite{neu}. One main feature is that $f_B$, the decay constant of
the $B$ meson entering Eq.~(\ref{eq:SR}) is calculated from a sum rule
itself\cite{SRfB}, which reduces the dependence of the resulting 
formfactors on the input parameters, in particular $m_b$, which is the
one-loop pole mass and taken to be $(4.80\pm0.05)\,$GeV. 
This procedure does not, however, reduce the formfactors' dependence on
the parameters describing the twist-2 DAs, which turns out to be rather
crucial. Despite much effort spent on both their calculation from
first principles and their extraction from experimental data, 
these so-called Gegenbauer
moments, $a_1$ (only for $K$), $a_2$ and $a_4$ (for all $P$) are not
known very precisely. 
Figure~\ref{fig:1} shows the dependence of $f_+^\pi(0)$ on
$a_2$ and $a_4$; the dots represent different determinations of these
parameters and illustrate the resulting spread in values of the formfactor. 
The situation is even more disadvantageous for the $K$, 
whose formfactors depend on the SU(3) breaking parameter $a_1^K$,
whose size and even sign are under
discussion\cite{SU(3)breaking}: at present, values as different as 
$-0.18$ and $+0.17$ (at $\mu=1\,{\rm GeV}$) are being quoted. 
Figure~\ref{fig:2} shows the
dependencies of (a) $f_+^K(0)$ and (b) $f_+^K(q^2)$ on this parameter;
evidently it is very important to determine its value more precisely.

Summarizing the detailed analysis of the uncertainties induced by both
external input and LCSR parameters, the final results for the
formfactors at zero momentum transfer obtained in Ref.\cite{neu} are:
$$
\renewcommand{\arraystretch}{1.2}\addtolength{\arraycolsep}{2pt}
\begin{array}[b]{lcl@{\qquad}lcl}
f_+^\pi(0) &=& 0.258\pm 0.031, & f_T^\pi(0)& =& 0.253\pm0.028,\\
f_+^K(0)& = &0.331\pm 0.041 + 0.25\delta_{a_1}, & f_T^K(0)& =&
0.358\pm0.037 + 0.31\delta_{a_1},\\
f_+^\eta(0)& =& 0.275\pm 0.036, & f_T^\eta(0)& =& 0.285\pm0.029.
\end{array}
\renewcommand{\arraystretch}{1}\addtolength{\arraycolsep}{-2pt}
$$
$\delta_{a_1}$ is defined as $a_1^K(1\,{\rm GeV}) - 0.17$, i.e.\ the
deviation of $a_1^K$ from the central value used in Ref.\cite{neu}.
For $f^{\pi,\eta}$ the total theoretical uncertainty ranges between
10\% to 13\%, for
$f^K$ it is 12\%, plus the uncertainty in $a_1$, which hopefully
will be clarified through an independent calculation in the not too
far future. The intrinsic, irreducible uncertainty of the sum rule
calculation is related to the dependence of the result on the sum rule
specific parameters $M^2$ and $s_0$ and estimated to be $\sim 7\%$.

\begin{figure}[tb]
$$
\begin{array}{@{}c@{\quad}c}
\epsfxsize=0.45\textwidth\epsffile{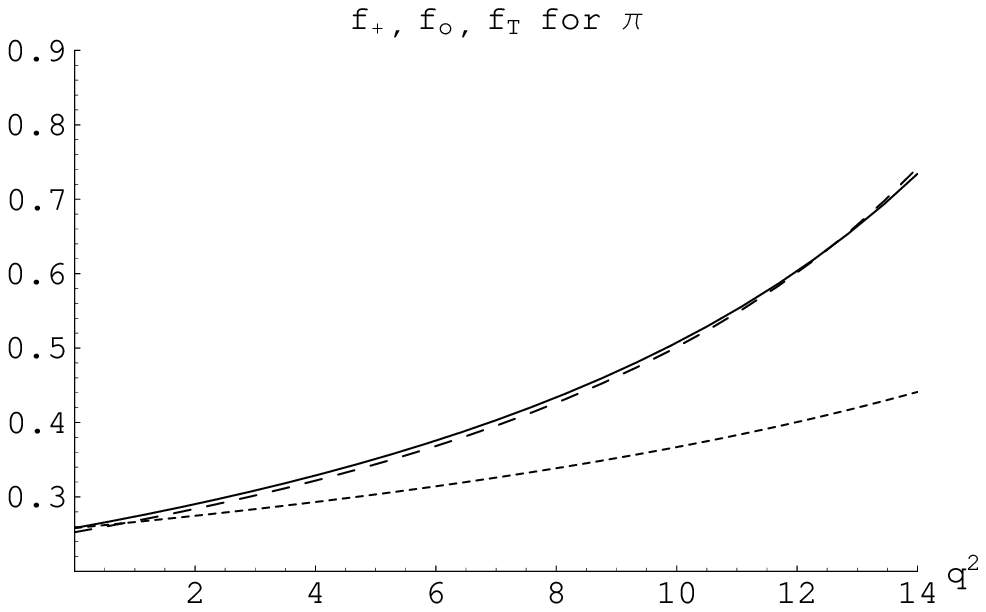} & 
\epsfxsize=0.45\textwidth\epsffile{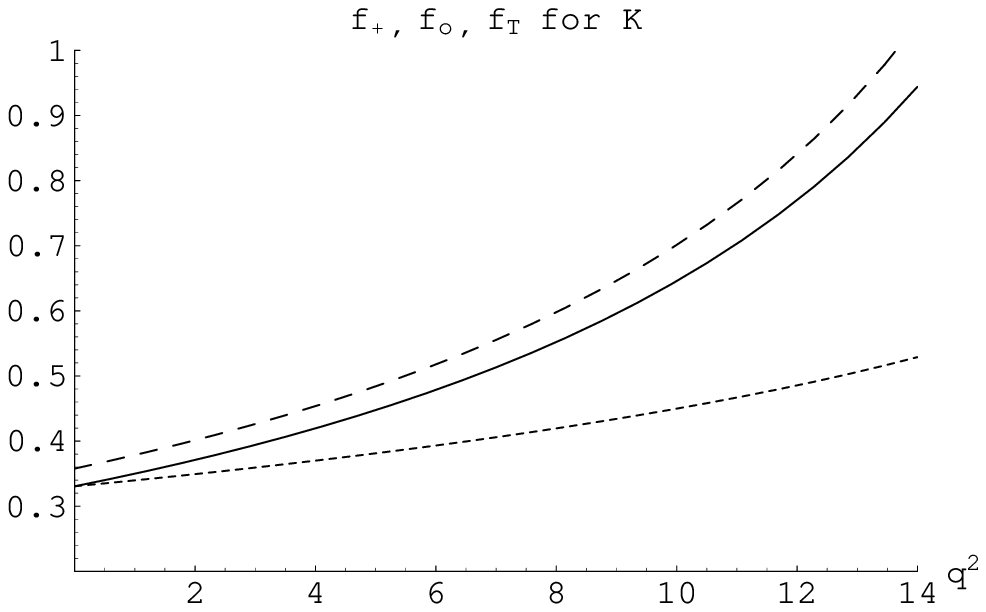}\\[5pt]
\epsfxsize=0.45\textwidth\epsffile{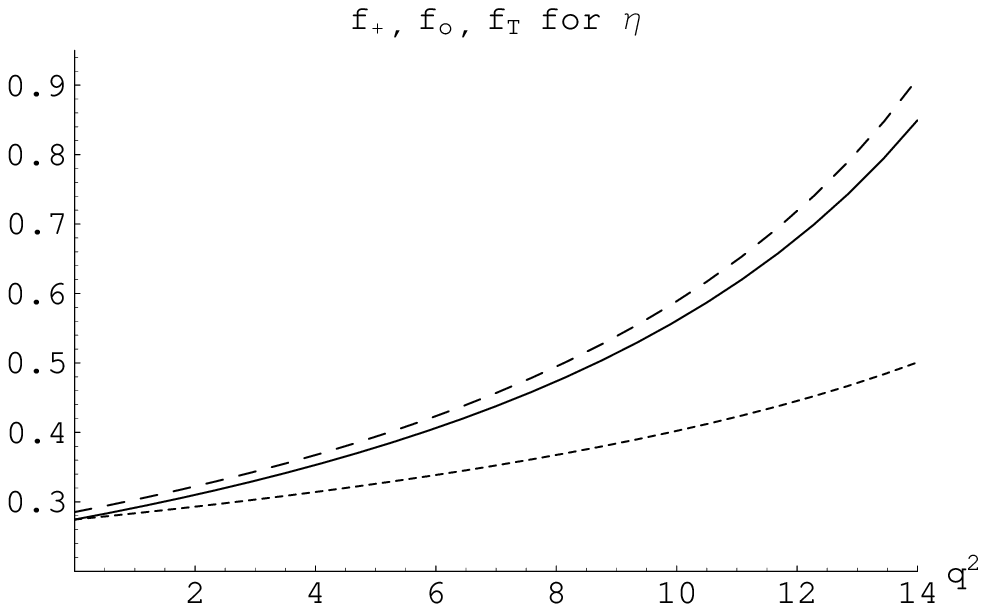}
\end{array}
$$
\caption[]{$f_+$ (solid lines), $f_0$ (short dashes) and $f_T$ (long
  dashes) as functions of $q^2$ for $\pi$, $K$
and $\eta$. The renormalisation scale of $f_T$ is chosen to be
$m_b$.}\label{fig:3}
\end{figure}

Turning to the $q^2$-dependence of formfactors,
it has to be recalled that LCSRs are only valid if the energy $E_P$ of the
final state meson, measured in the rest frame of the decaying $B$, is
large, i.e.\ if $q^2=m_B^2-2 m_B E_P$ is not too large; specifically, we
choose $E_P>1.3\,$GeV, i.e.\ $q^2\leq 14\,{\rm GeV}^2$. The resulting 
formfactors are plotted in Fig.~\ref{fig:3}, using central values for
the input parameters. 
In order to allow a simple implementation of these results in actual
applications, and also in order to provide predictions for the full physical
regime $0\leq q^2\leq (m_B-m_P)^2\approx 25\,{\rm GeV}^2$, it is
necessary to find parametrizations of $f(q^2)$ that 
\begin{itemize}
\item reproduce the data below $14\,{\rm GeV}^2$ with good accuracy; 
\item provide an extrapolation to $q^2>14\,{\rm GeV}^2$ that is 
consistent with the expected analytical properties of the formfactors
and reproduces the lowest-lying resonance
(pole) with $J^P=1^-$ for  $f_+$ and $f_T$.\footnote{For $f_0$, the
lowest pole with quantum numbers $0^+$ lies above the two-particle threshold
starting at $(m_B+m_P)^2$ and hence is not expected to feature prominently.} 
\end{itemize}
As shown in Ref.\cite{neu}, the following parametrizations are appropriate:
\begin{itemize}
\item for $f_{+,T}^\pi$:
\begin{equation}
\label{eq:doubledouble}
f(q^2) = \frac{r_1}{1-q^2/m_1^2} + \frac{r_2}{1-q^2/m_{\rm fit}^2} \quad ,
\end{equation}
where $m_1^\pi$ is the mass of $B^*(1^-)$, $m_1^\pi=5.32\,$GeV; the
fit parameters are $r_1$, $r_2$ and $m_{\rm fit}$;
\item for $f_{+,T}^{K,\eta}$:
\begin{equation}
\label{eq:Keta}
f(q^2) = \frac{r_1}{1-q^2/m_{1}^2} + \frac{r_2}{(1-q^2/m_{1}^2)^2},
\end{equation}
where $m_1$ is the mass of the $1^-$ meson in the corresponding
channel, i.e.\ 5.32~GeV for $\eta$ and 5.41~GeV for $K$; 
the fit parameters are $r_1$ and $r_2$; 
\item for $f_0$:
\begin{equation}
\label{eq:singlesingle}
f_0(q^2) = \frac{r_2}{1-q^2/m_{\rm fit}^2}\,,
\end{equation}
the fit parameters are $r_2$ and $m_{\rm fit}$.
\end{itemize}
The central results for the fit parameters are collected in
Tab.~\ref{tab:fitpars}.
\begin{table}[tb]
\tbl{Fit parameters for $f(q^2)$.
$m_1$ is the vector meson mass in the corresponding channel:
$m_1^{\pi,\eta} = m_{B^*} = 5.32\,$GeV and $m_1^K = m_{B^*_s} = 5.41\,$GeV. 
The scale of $f_T$ is $\mu = 4.8\,$GeV.\label{tab:fitpars}}
{\begin{tabular}{c|rrcr}
& $r_1$ & $r_2$ & $(m_1)^2$ & $m_{\rm fit}^2$\\\hline
$f_+^{\pi}$ &0.744 & -0.486 & $(m_{1}^\pi)^2$ & 40.73\\
$f_0^{\pi}$ & 0 & 0.258 & - & 33.81\\
$f_T^{\pi}$ & 1.387 & -1.134 & $(m_{1}^\pi)^2$ & 32.22\\
$f_+^{K}$ & 0.162 & 0.173 & $(m_1^K)^2$ & -\\
$f_0^{K}$ & 0 & 0.330 & - & 37.46\\
$f_T^{K}$ & 0.161 & 0.198 & $(m_1^K)^2$ & -\\
$f_+^{\eta}$ & 0.122 & 0.155 & $(m_1^\eta)^2$ & -\\
$f_0^{\eta}$ & 0 &0.273  & - & 31.03\\
$f_T^{\eta}$ & 0.111 & 0.175 & $(m_1^\eta)^2$ & -\\
\hline
\end{tabular}
}
\end{table}
The quality of all fits is very good and the maximum deviation between
LCSR and fitted result is 2\% or better. The impact of the {\it
  extrapolation} of the fit formulas to $q^2>14\,{\rm GeV}^2$ is of
phenomenological relevance mainly for $B\to\pi e \nu$, relevant for
the determination of $|V_{ub}|$ from experiment. We have estimated the
effect of the extrapolation on the decay rate by implementing
different parametrisations for $f_+^\pi$, which all fit the LCSR
result very well for $q^2<14\,{\rm GeV}^2$, but differ for larger
$q^2$, the main distinguishing feature being the positions of the
poles. We find that for reasonable parametrisations, that is such that
do not exhibit too strong a singularity at $q^2=m_1^2$, the total
rates differ by not more than 5\%, the difference becoming smaller if
an cut-off on the maximum invariant mass of the lepton pair is
implemented, which implies that the extrapolation is well under control.

\section{Summary \& Conclusions}

LCSRs provide accurate results for weak decay formfactors of the $B$
meson into light mesons, in particular $\pi$, $K$ and $\eta$. The
results depend on sum rule specific input parameters which generate an
irreducible ``systematic'' uncertainty of the approach 
estimated to be $\sim 7\%$. Additional uncertainties are induced by
imprecisely known hadronic input parameters, in particular the
Gegenbauer moments $a_{1,2,4}$ describing the leading-twist
light-meson distribution amplitudes. An improved determination of
these parameters would be very welcome. The present total uncertainty
of the formfactors at zero momentum transfer varies between 10 and
13\%, but becomes smaller at larger $q^2$. LCSR
calculations require the energy of the final state meson to be large
in the rest-frame of the decaying $B$ and hence are valid only for 
not too large momentum transfer $q^2$; the maximum eligible $q^2$ is
chosen to be $14\,{\rm GeV}^2$. The $q^2$-dependence of the
formfactors can be cast into simple parametrizations in terms of two
or three parameters, which also capture the main features of the
analytical structure and are expected to be valid in the full
kinematical regime $0\leq q^2\leq (m_B-m_P)^2$. The total uncertainty
introduced by the extrapolation of the formfactors to $q^2$ larger
than the sum rule cut-off $14\,{\rm GeV}^2$ is
estimated to be $\sim 5\%$ for the semileptonic rate $\Gamma(B\to\pi e
\nu)$.

Ref.\cite{neu} also contains a detailed breakdown of the dependence of
the formfactors on the Gegenbauer moments, which not only allows one
to recalculate the formfactors once these parameters are
determined more precisely, but also makes it possible to consistently
assess their impact on nonleptonic decay amplitudes (e.g.\
$B\to\pi\pi$) treated in QCD factorisation.

The LCSR approach is complementary to standard lattice calculations, in the
sense that it works best for large energies of the final state meson
(i.e.\ small $q^2$), whereas lattice calculations work best for small
energies -- a situation that may change in the
future with the implementation of moving NRQCD\cite{davies}.
Previously, the LCSR results for $f_{+,0}^\pi$ at small and 
moderate $q^2$ were found to 
nicely match\cite{match} the lattice results obtained for large
$q^2$. 
The situation will have to be reassessed in view of our
new results and it will be very interesting to see if and how it will 
develop with 
further progress in both lattice and LCSR calculations.

\end{document}